\documentclass[conference]{IEEEtran}
\IEEEoverridecommandlockouts
\usepackage{cite}
\usepackage{amsmath,amssymb,amsfonts}
\usepackage{algorithmic}
\usepackage{graphicx}
\usepackage{textcomp}
\usepackage{xcolor}
\usepackage{amsmath}
\usepackage{fullpage}
\usepackage{times}
\usepackage{fancyhdr,graphicx,amsmath,amssymb}
\usepackage[ruled,vlined]{algorithm2e}       
\usepackage{subcaption}
\DeclareMathOperator*{\argmin}{arg\,min}

\def\BibTeX{{\rm B\kern-.05em{\sc i\kern-.025em b}\kern-.08em
    T\kern-.1667em\lower.7ex\hbox{E}\kern-.125emX}}
\begin{document}

\title{Non-Orthogonal Time-Frequency Space Modulation\\
}

\author{\IEEEauthorblockN{1\textsuperscript{st} Mahdi Shamsi}
	\IEEEauthorblockA{\textit{EE. dept. of Sharif Uni. of Tech.} \\
		\textit{Advanced Comm. Research Institute (ACRI).}\\
		\textit{Multimedia and Signal processing Lab. (MSL).}\\
		Mahdi.Shamsi@alum.sharif.edu}
	\and
	\IEEEauthorblockN{2\textsuperscript{nd} Farokh Marvasti}
	\IEEEauthorblockA{\textit{EE. dept. of Sharif Uni. of Tech.} \\
		\textit{Advanced Comm. Research Institute (ACRI).}\\
		\textit{Multimedia and Signal processing Lab. (MSL).}\\
		marvasti@sharif.edu}
	}
\maketitle

\begin{abstract}
This paper proposes a Time-Frequency Space Transformation (TFST) to derive non-orthogonal bases for modulation techniques over the delay-doppler plane. A family of Overloaded Delay-Doppler Modulation (ODDM) techniques is proposed based on the TFST, which enhances flexibility and efficiency by expressing modulated signals as a linear combination of basis signals. A Non-Orthogonal Time-Frequency Space (NOTFS) digital modulation is derived for the proposed ODDM techniques, and simulations show that they offer high-mobility communication systems with improved spectral efficiency and low latency, particularly in challenging scenarios such as high overloading factors and Additive White Gaussian Noise (AWGN) channels. A modified sphere decoding algorithm is also presented to efficiently decode the received signal. The proposed modulation and decoding techniques contribute to the advancement of non-orthogonal approaches in the next-generation of mobile communication systems, delivering superior spectral efficiency and low latency, and offering a promising solution towards the development of efficient high-mobility communication systems.
\end{abstract}


\section{Introduction}
In the forthcoming evolution of mobile communication systems, notable challenges arise from channel impairments, particularly within Doppler channels. This phenomenon is also accentuated in mobile sensor networks, where such communication distortions possess the potential to propagate errors, thereby exerting a substantial influence on the overall network performance \cite{shamsi2021flexible,shamsi2023distributed}.

To fill this need, Orthogonal Time Frequency Signaling (OTFS) was proposed, which compensates for channel impairments \cite{monk2016otfs}. Recent studies have demonstrated that OTFS can achieve the same spectral efficiency performance as Orthogonal Frequency Division Multiplexing (OFDM) based techniques. Moreover, OTFS can be utilized in high-mobility user scenarios, which is one of the proposed goals of the next mobile generation according to 3GPP visions.
However, despite its performance comparable to OFDM, the 2-d kernels of OTFS result in inevitable, larger latencies during communication procedures. To address these shortcomings, we propose a new modulation technique that retains the advantages of OTFS and OFDM but omits the orthogonality.
Our approach introduces a Time-Frequency Space Transformation (TFST) to derive non-orthogonal bases and create a class of modulation techniques over the delay-doppler plane. The class includes previously studied techniques, such as Time Division Multiplexing (TDM), Frequency Division Multiplexing (FDM), Code Division Multiple Access (CDMA), OFDM, and OTFS.

While researchers have studied Faster Than Nyquist (FTN) signaling \cite{mazo1975faster} and overloaded CDMA \cite{alishahi2011design} to address these constraints, an additional proposed solution is Spectrally Efficient FDM (SEFDM)\cite{xu2013improved, shamsi2023enhancing}, which has demonstrated promising results in increasing spectral efficiency without sacrificing signal quality. However, like FTN signaling, overloaded CDMA and SEFDM have not yet been widely accepted beyond research. Our proposed modulation technique improves upon these existing methods and provides a more efficient and effective solution for high-mobility communication systems.
This new technique has the potential to become the standard for high-mobility communication systems such as 6G and beyond, where low latency and high spectral efficiency are vital.
In summary, our key contributions encompass:

\begin{itemize}
	\item Introduction of a new 2-d signal representation and its properties,
	\item Derivation of a new class of 2-d modulations,
	\item Reduction of detection complexity  through the proposal of 2-d Sphere Decoding.
\end{itemize}

In Section \ref{sect:overload}, we propose the TFST and derive a new class of Delay-Doppler (DD) modulation techniques. These modulation techniques offer the benefits of both OTFS and OFDM, without the orthogonality constraints. Section \ref{sect:SD_IM} is dedicated to introducing a 2-d version of Sphere Decoding (SD) to improve the performance of the proposed approach. In Section \ref{sect:sim}, we showcase the performance of the proposed technique using simulations. Finally, Section \ref{sect:con} concludes the paper by summarizing the advantages of the proposed modulation technique and its potential to become the standard for high-mobility communication systems in the future.
\section{Overloading Delay-Doppler Modulation Techniques}
\label{sect:overload}
This section introduces a novel 2-d modulation class facilitated by the newly proposed Time-Frequency Space Transformation (TFST). TFST allows the analysis of complex continuous time signals in a 2-d format, specifically in the delay-Doppler domain. Key properties, such as shift invariance and direct connections to time and Fourier signal representations, are examined.

Utilizing the TFST, we derive a group of signal bases that effectively span the domain, narrowing them down to specific time and frequency ranges. This results in a set of bases for the proposed modulation techniques, offering superior flexibility and efficiency compared to traditional methods by providing a comprehensive signal representation in a 2-d framework.

The TFST of an arbitrary complex-valued signal $x(t)$ is defined as ($-\infty < \tau < \infty \,,\, -\infty < \nu < \infty$):
\begin{eqnarray}
\label{mzdef1}
{\mathcal M}^{\lambda,\mu}_x(\tau, \nu) \triangleq \sqrt{\lambda T} \,\sum\limits_{n=-\infty}^{+\infty} \, x(\tau + n\lambda T) \, e^{-j 2 \pi \frac{n \nu T}{\mu} },\notag
\end{eqnarray}
where $\tau$ represents the delay parameter, $\nu$ the Doppler frequency parameter, and $(\lambda,\mu)$ the transform parameters. 

Shift invariance: The shift invariance property of the TFST can be shown by considering a signal that has undergone a delay and a Doppler shift. Let $r(t) = x(t - \tau_0) e^{j 2 \pi \nu_0 (t - \tau_0)}$, where $\tau_0$ and $\nu_0$ represent the delay and Doppler shift parameters, respectively. 
The TFST of $r(t)$ can be computed as:
\begin{eqnarray}
\label{mZakmod2}
{\mathcal M}^{\lambda,\mu}_r(\tau, \nu) =  {\mathcal M}^{\lambda,\mu}_x(\tau-\tau_0, \nu-\lambda\mu \nu_0) e^{-j 2 \pi\nu_0 (\tau - \tau_0) }\notag.
\end{eqnarray}

Periodicity: Another important property of the TFST is its periodicity in both the time and frequency domains. Specifically, for any arbitrary complex-valued signal $x(t)$, we have:
\begin{eqnarray}
{\mathcal M}^{\lambda,\mu}_x(\tau + \lambda T, \nu)  &=&  e^{j 2 \pi \nu T/\mu} \, {\mathcal M}^{\lambda,\mu}_x(\tau,\nu),\notag\\
{\mathcal M}^{\lambda,\mu}_x \left(\tau,\nu + \mu \Delta f \right) &=& {\mathcal M}^{\lambda,\mu}_x(\tau,\nu),\notag
\end{eqnarray}
where $\Delta f = 1/T$. Together, the shift invariance and periodicity properties of the TFST enable efficient analysis and processing of time-varying signals in the delay-Doppler domain.

Multiplication property: The TFST of the product of two signals $a(t)$ and $b(t)$, denoted $c(t) = a(t) \times b(t)$, is given by ${\mathcal M}^{\lambda,\mu}_c(\tau, \nu) =  \frac{1}{\lambda\mu}\sqrt{\lambda T} \times I$, where $I$ is defined as
\begin{eqnarray}
\label{zeqn13}
I& \triangleq&\int\limits_{0}^{\mu\Delta f} {\mathcal M}^{\lambda,\mu}_a(\tau,\nu - \nu') {\mathcal M}^{\lambda,\mu}_b(\tau , \nu') d\nu' \nonumber \\
& =&  \frac{1}{\lambda\mu}\sqrt{\lambda T} \int\limits_{0}^{\mu\Delta f} {\mathcal M}^{\lambda,\mu}_a(\tau,\nu') {\mathcal M}^{\lambda,\mu}_b(\tau, \nu - \nu') d\nu'.\nonumber
\end{eqnarray}

Convolution property: The TFST of the convolution of two signals $a(t)$ and $b(t)$, denoted $c(t) = a(t) \star b(t)$, is given by
\begin{eqnarray}
\label{zeqn12}
{\mathcal M}^{\lambda,\mu}_c(\tau, \nu) =  \frac{1}{\sqrt{\lambda T}} \int\limits_{0}^{\lambda T}  {\mathcal M}^{\lambda,\mu}_a(\tau - \tau',\nu) \, {\mathcal M}^{\lambda,\mu}_b(\tau' , \nu) \, d\tau' \nonumber \\
=  \frac{1}{\sqrt{\lambda T}} \int\limits_{0}^{\lambda T} {\mathcal M}^{\lambda,\mu}_a(\tau',\nu) \,{\mathcal M}^{\lambda,\mu}_b(\tau - \tau' , \nu) \, d\tau'.\notag
\end{eqnarray}

The time domain signal $x(t)$ and its Fourier transform ${\mathcal F}_x(f) = \int\limits_{-\infty}^{\infty} x(t) e^{-j 2 \pi f t} \, dt$ can be obtained from its TFST representation ${\mathcal M}_x(\tau,\nu)$ using the following equations:
\begin{eqnarray}
\label{mzeqn6}
x(t) & = & \frac{1}{\lambda\mu}\sqrt{ \lambda T}\int\limits_{0}^{\mu\Delta f} \, {\mathcal M}_x^{\lambda,\mu}(t, \nu) \, d\nu, \notag\\
\label{mzeqn7}
{\mathcal F}_x(f) & = & \frac{1}{\sqrt{\lambda T}} \int\limits_{0}^{\lambda T} {\mathcal M}_x^{\lambda,\mu}(\tau, \lambda\mu f) \, e^{-j 2 \pi f \tau} \, d\tau.\notag
\end{eqnarray}

\subsection{Derivation of the Modulation Technique}
A signal in the delay domain, located at $\tau_0$ and in the Doppler domain, located at $\nu_0$ ($0 \leq \tau_0 < T$, $0 \leq \nu_0 < \Delta f$), can be expressed as:
\begin{eqnarray}
\label{lceqn1}
{\mathcal M}_{(p, \tau_0, \nu_0)}^{\lambda,\mu}(\tau, \nu) \triangleq \hspace{-2mm}
\sum\limits_{m=-\infty}^{+\infty} \sum\limits_{n = -\infty}^{+\infty}  {\Big (} e^{j 2 \pi \nu_0 n \mu^{-1} T} \nonumber \\ \delta(\tau - \tau_0 - n\lambda T)\delta(\nu - \nu_0 - m \mu\Delta f)  {\Big )}
\end{eqnarray}
The time domain signal $p_{(\tau_0, \nu_0)}(t)$, with a DD representation ${\mathcal M}_{(p, \tau_0, \nu_0)}(\tau, \nu)$ as shown in (\ref{lceqn1}), can be expressed as:
\begin{eqnarray}
\label{plemeqn2}
p_{(\tau_0, \nu_0)}^{\lambda,\mu}(t)\hspace{-1mm} =\hspace{-1mm} \frac{1}{\lambda\mu} \sqrt{\lambda T} \sum\limits_{n=-\infty}^{+\infty} e^{j 2 \pi \nu_0 n \mu^{-1}  T } \delta(t  - \tau_0 - n\lambda T).\notag
\end{eqnarray}
It can be shown that the time domain signals $p_{(\tau_0, \nu_0)}(t)$, where $0 \leq \tau_0 < \lambda T$ and $0 \leq \nu_0 < \mu\Delta f$, form a basis for the space of time domain signals.
Any time domain signal $x(t)$ can be expressed as a linear combination of the basis signals $p_{(\tau_0, \nu_0)}(t)$, i.e.:
\begin{eqnarray}
\label{thmbasiseqn1m}
x(t) = \int_{0}^{\lambda T} \int_{0}^{\mu \Delta f} c_x^{\lambda,\mu}(\tau_0, \nu_0) \, p_{(\tau_0, \nu_0)}^{\lambda,\mu}(t) \, d\tau_0 \, d\nu_0, \nonumber \\
c_x^{\lambda,\mu}(\tau_0, \nu_0) = \int_{-\infty}^{+\infty}{ p^{\lambda,\mu}_{(\tau_0, \nu_0)}(t)}^* \, x(t) \, dt.\notag
\end{eqnarray}
And the coefficient $c_x(\tau_0, \nu_0)$ corresponding to the basis signal $p_{(\tau_0, \nu_0)}(t)$ is the value of the TFST representation of $x(t)$ at $\tau = \tau_0$ and $\nu = \nu_0$, i.e.:
\begin{eqnarray}
\label{prfceqn}
c_x^{\lambda,\mu}(\tau_0, \nu_0) = \frac{1}{\lambda\mu} {\mathcal M}^{\lambda,\mu}_x(\tau_0,\nu_0).\notag
\end{eqnarray}
\subsection{Non-Orthogonal Time Frequency Space}
The derivation of Non-Orthogonal Time Frequency Space (NOTFS) modulation begins by defining the basis signals $\psi^{(q,s),(\lambda,\mu)}_{(\tau_0, \nu_0)}(t)$ as a product of the time and frequency pulses $q(t)$ and $S(f)$, and the TD signal $p_{(\tau_0, \nu_0)}^{\lambda,\mu}(t)$, as follows: 
\begin{eqnarray}
\label{basiseqn1}
\psi^{(q,s),(\lambda,\mu)}_{(\tau_0, \nu_0)}(t)  \triangleq  \left( p_{(\tau_0, \nu_0)}^{\lambda,\mu}(t) \, q(t) \right)\star  s(t),
\begin{cases}
 0 \leq \tau_0 < \lambda T \nonumber\\
 0 \leq \nu_0 < \mu \Delta f
\end{cases}
\end{eqnarray}
where $q(t) \approx  0 \,,\, t \notin [0 \,,\, N\epsilon T)$ and
$\vert {\mathcal F}_s(f) \vert \hspace{-3mm}  = \hspace{-3mm} \left\vert \int_{-\infty}^{+\infty} \hspace{-2mm} s(t) e^{-j 2 \pi f t} \, dt \right\vert \, \approx \, 0 \,,\, f \notin [0 \,,\, M \kappa \Delta f)$. When rectangular pulses are used, a simplified expression for the basis signals is obtained as:
\begin{eqnarray}
\label{basisqeqn}
 \hspace{-0mm}\psi^{(q,s),(\lambda,\mu)}_{(\tau_0, \nu_0)}(t)  = \frac{\sqrt{\lambda T}}{\lambda\mu}  \sum\limits_{n=0}^{N'-1} e^{j 2 \pi \nu_0 n \mu^{-1} T} \, s(t - \tau_0 - n\lambda T),\notag
\end{eqnarray}
where $N'\approx  \frac{\epsilon}{\lambda}\times N$.
Applying the TFST representation to the basis signals leads to 
\begin{eqnarray}
\label{eqn43}
{\mathcal M}^{\lambda,\mu}_{\psi, \tau_0, \nu_0}(\tau, \nu) = \frac{1}{\lambda\mu}{\mathcal M}^{\lambda,\mu}_q\left( \tau_0 , \nu - \nu_0 \right) \, {\mathcal M}^{\lambda,\mu}_s\left( \tau - \tau_0, \nu \right),\notag
\end{eqnarray}
which relates the TFST representation of the basis signals to that of the pulses $q(t)$ and $s(t)$.

Now we can perform a concentration analysis to determine the magnitude of an arbitrary modulation base. Through this analysis, as shown in
\begin{eqnarray}
 \left\vert  {\mathcal M}^{\lambda,\mu}_{\psi, \tau_0, \nu_0}(\tau, \nu) \right\vert^2 = \frac{1}{(\lambda\mu)^2}\frac{\sin^2\left( \pi N''\frac{ (\nu - \nu_0) }{\mu \Delta f}\right)}{\sin^2\left( \pi \frac{ (\nu - \nu_0) }{\mu \Delta f} \right)} \times\dots\notag\\
\, \frac{\sin^2 \left( \pi M'\frac{ (\tau - \tau_0)}{\lambda T} \right) }{\sin^2 \left( \pi \frac{ (\tau - \tau_0)}{\lambda T} \right) }, \nonumber
\end{eqnarray}
we can observe that the signal is concentrated around $(\tau_0, \nu_0)$.
With this in mind, each modulation base can be defined by the following equation:
\begin{eqnarray}
\label{eqn27}
\chi^{\lambda,\mu}_{(k,l)}(t)  &\triangleq & \frac{1}{\sqrt{MN}} \psi^{(q,s),(\lambda,\mu)}_{\left( \tau_0 = \frac{l\phi T}{M}, \nu_0	= \frac{k \theta \Delta f }{N} \right)}(t).\notag
\end{eqnarray} 
This suggests an Overloaded Delay-Doppler Modulation (ODDM) where the modulated signal is expressed as:
\begin{eqnarray}
x(t) & \hspace{-3mm} = & \hspace{-3mm} \sum\limits_{k=0}^{N-1}\sum\limits_{l=0}^{M-1} x[k,l] \,\chi^{\lambda,\mu}_{(k,l)}(t).\notag
\end{eqnarray}

By digitally implementing the ODDM, the group of NOTFS modulations can be shown as:

\begin{eqnarray}
x(t)  &   \approx  &  \frac{\rho}{\mu} \frac{1}{\sqrt{MN}} \sum\limits_{n=0}^{N''-1} \sum\limits_{m=0}^{M''-1} \hspace{-1mm} g(t - n\lambda T)\times\dots\notag\\
&&X_{\mbox{\tiny{{TF}}}}[n,m] \, e^{j 2 \pi m  \rho\Delta f (t - n\lambda T)}, \nonumber \\
X_{\mbox{\tiny{{TF}}}}[n,m] & \hspace{-2mm}  \triangleq & \hspace{-2mm} \sum\limits_{k=0}^{N''-1}\sum\limits_{l=0}^{M''-1} x[k,l] \, e^{j 2 \pi \left( \frac{\theta}{\mu}\frac{n k}{N} - \phi \rho\frac{m l}{M} \right)}, \nonumber \\
& & \begin{cases}
 n=0,1,\cdots, N''-1 \\
m=0,1,\cdots, M''-1\notag
\end{cases}, \nonumber \\
g(t) &  \hspace{-2mm} \triangleq &  \hspace{-2mm} \begin{cases}
\frac{1}{\sqrt{\lambda T}}  \,,\, t \in [0 \,,\, \lambda T)  \\
0  \,\,\,,\, \mbox{\small{{otherwise}}}.\notag
\end{cases}.
\end{eqnarray}
For the rest of the article, we will focus on two typical groups of essentially equal parameters based on overloading parameters ($\alpha$ and $\beta$ control compression in the time and frequency domains, respectively), and continue with the latter:

Group 1: $\theta=\mu,\Phi.\rho = 1,\kappa=\rho=\beta,\, \lambda=\epsilon=\alpha $

Group 2: $\lambda=\rho=\kappa=\epsilon=1,\,\frac{\theta}{\mu}=\alpha,\,\Phi = \beta$.
\section{Sphere Decoding and Inverse Systems}
\label{sect:SD_IM}
Since the ODDM techniques are two-dimensional in nature, we suggest a modified version of the SD algorithm. To establish a suitable benchmark for comparison, we also offer a brief overview of an iterative inverse methods family that can serve as an initial value for the decoding process.
\begin{figure*}[h]
	{\includegraphics[width=\textwidth,height=.5\textwidth]{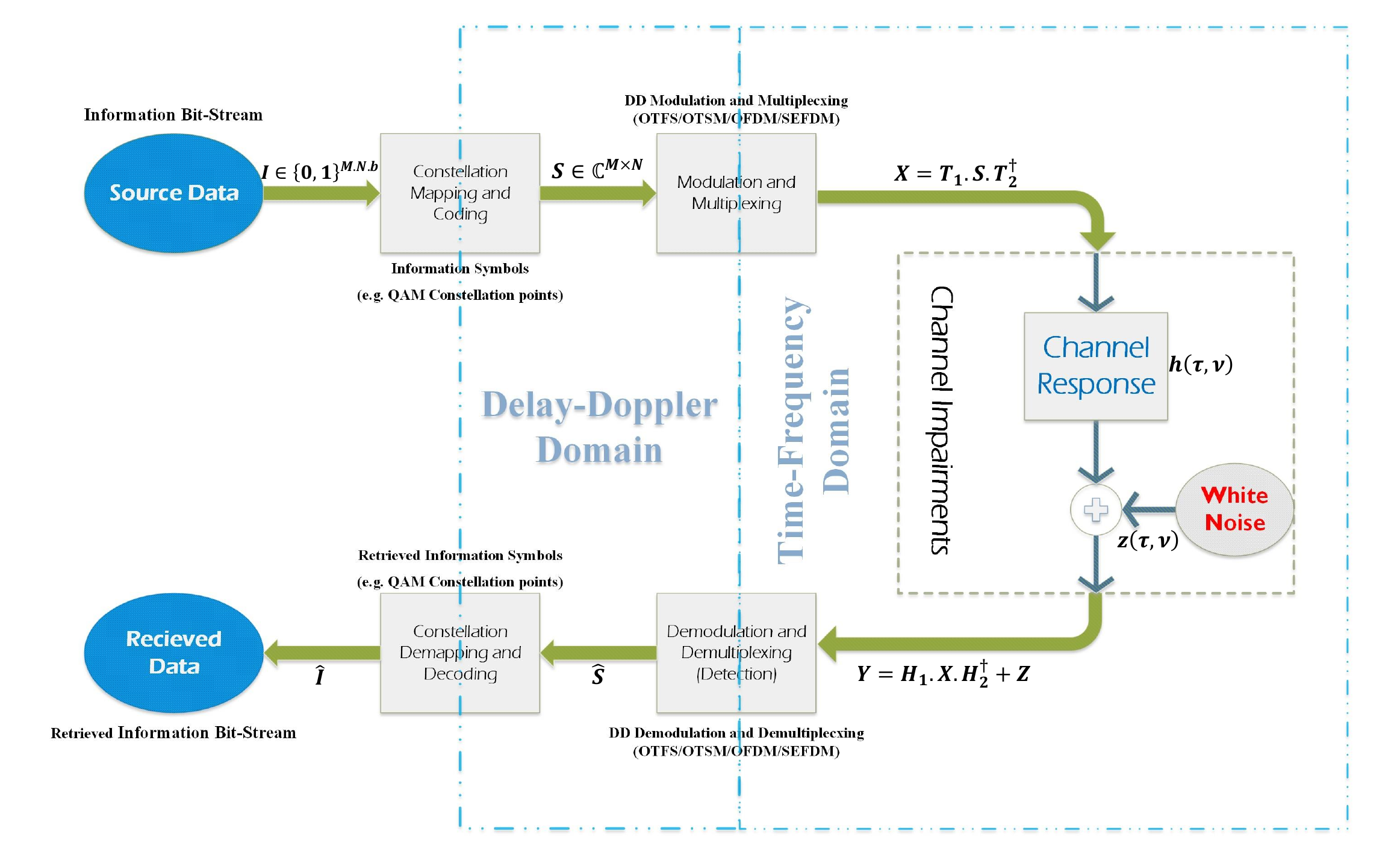}}
	\caption{Typical block-diagram of Delay-Doppler (DD) modulation techniques.}
	\label{fig:BLK}
\end{figure*}
\subsection{2-D Sphere Decoding}
To describe our communication system depicted in Fig. \ref{fig:BLK}, we use $\boldsymbol X$ as the transmitted signal. We represent the received signal frame by $\boldsymbol Y = H_1.\boldsymbol X.H_2^\dagger + \boldsymbol Z$, where  $\boldsymbol Z$ is the additive noise of the communication channel at the receiver.
For the sake of brevity, we use the notation $(G,H)$. Specifically, we express $G$ as $G = T_1^\dagger .H_1.T_1$ and $H$ as $H = T_2^\dagger .H_2.T_2$, define the objective function as $J(\boldsymbol S)\triangleq\left |\left|\boldsymbol Y_T-G\boldsymbol SH^\dagger\right |\right|^2_F$ (where $\boldsymbol{Y_T}\triangleq T_1.\boldsymbol Y.T_2^\dagger$), and set the goal to solve $\widehat{\boldsymbol S} = \argmin_{ \boldsymbol S \in \mathbb{A}^{M\times N}} \,  J(\boldsymbol S)<g^2$. Here, $\mathbb{A}$ is the set of constellation points of sending symbols, and $g$ is the search radius. To solve this problem, the SD algorithm presented in Alg.\ref{alg:2-d_SD} (with update routine $\Psi$ as in Alg.\ref{alg:update 2-d_SD}) is used.

We first calculate the QR decompositions of $H$ and $G$, denoted as $G=Q_GR_G$ and $H=Q_HR_H$. Accordingly, the partial objective functions are defined using $R=R_G$, $L =R_H^\dagger$, and $\boldsymbol U = Q_G^\dagger.\boldsymbol Y_T.Q_H$, as follows\footnote{the indexing method for a matrix (or a frame) $A$ is defined as:
	$A_{m:M, n:N}\triangleq\begin{bmatrix}
	A_{m,n} & \dots & A_{m,N} \\ 
	\vdots  &  \ddots &\vdots  \\ 
	A_{M,n} & \dots & A_{M,N}
	\end{bmatrix},
	A_{m:M, n}\triangleq\begin{bmatrix}
	A_{m,n} \\ 
	\vdots  \\ 
	A_{M,n}
	\end{bmatrix}$, and $A_{m, n:N}\triangleq\begin{bmatrix}
	A_{m,n} & \dots & A_{m,N}
	\end{bmatrix}$.}:
\begin{eqnarray}
J_{m,n}(S_{m:M, n:N})\triangleq \left |U_{m,n}-R_{m, m:M}S_{m:M, n:N}L_{m:M, n}\right |^2.\notag
\end{eqnarray} 
Thus, the overall objective function can be rewritten as $J(\boldsymbol S)\triangleq\sum_{m=0}^{M}\sum_{n=0}^{N}J_{m,n}(S_{m:M, n:N})$, which facilitates applying the SD algorithm to solve for the optimal solution $\widehat{\boldsymbol S}$.

It is worth mentioning that the use of 2-D SD can significantly reduce computational complexity compared to 1D SD, as demonstrated in Table \ref{table:complexity2}. This complexity reduction is crucial in effectively optimizing communication systems with large numbers of symbols or frames. Consequently, the 2-D SD algorithm improves system performance while simultaneously mitigating errors.
\begin{algorithm}
	\footnotesize{\SetAlgoLined
	\caption{2-D Sphere Decoding.}
	\KwResult{$\hat{S}=arg\min\limits_{\boldmath S} J(\boldmath S)\, s.t. S_{m,n}\in\, \mathbb{A}$}
	Initialization: $\hat{X} \in\{0\}^{M\times N\times \kappa }, \hat{J}\in\{0\}^{\kappa}$\\
	Update: $\hat{X},\hat{J} \leftarrow \Psi (\hat{X},\hat{J}, M,N)$\\
	\For{$i = 1: \min(M,N)-1$}{
		\For{$k = 1: i-1$}{
			Update: $\hat{X},\hat{J} \leftarrow \Psi (\hat{X},\hat{J}, M-k,N-i)$\\
			Update: $\hat{X},\hat{J} \leftarrow \Psi (\hat{X},\hat{J}, M-i,N-k)$
		}
		Update: $\hat{X},\hat{J} \leftarrow \Psi (\hat{X},\hat{J}, M-i,N-i)$
	}
	\uIf{$M>N$}{%
		\For{$i = N: M-1$}{
			\For{$k = 0 : N-1$}{
				Update: $\hat{X},\hat{J} \leftarrow \Psi (\hat{X},\hat{J},M-i,N-k)$
			}
		}
	}\ElseIf{$M<N$}{
		\For{$i = M: N-1$}{
			\For{$k = 0 : M-1$}{
				Update: $\hat{X},\hat{J} \leftarrow \Psi (\hat{X},\hat{J},M-k,N-i)$
			}
		}
	}
	\Return $\hat{S}=\hat{X}_{:,:,1}$
	\label{alg:2-d_SD}}
\end{algorithm}
\begin{algorithm}
	\footnotesize{\SetAlgoLined
	\caption{Updating Procedure of the 2-D SD.}
	\KwResult{$\hat{X},\hat{J}\leftarrow X\text{'s with least}\; J(X)$;\\ least $\kappa$ estimations}
	Input: Estimations ($\hat{X}$), Loss values ($\hat{J} $)\\
	Initialization: $Y\leftarrow \hat{X}$\\
	\For{each $\hat{J}_i\leq g^2$}{
		\For{each $s_l$ in constellation}{
			$\hat{X}_{m,n,i}\leftarrow s_l$\\
			$J^{\text{temp}}(i,l)\leftarrow\hat{J}_i+J_{m,n}(\hat{X}_{:,:,i})$
		}
		$I,P \leftarrow $ indices[sort $J^{\text{temp}}$, ascending order]
	}
	\For{$t = 1,\dots,\kappa$}{
		$\hat{J}_t\leftarrow J^{\text{temp}}(I_t,P_t)$\\
		$\hat{X}_{:,:,t}\leftarrow Y_{:,:,I_t}$\\
		$\hat{X}_{m,n,t}\leftarrow s_{P_t}$
	}
	\Return $\hat{X},\hat{J}$
	\label{alg:update 2-d_SD}}
\end{algorithm}

\begin{table}[h]
	\caption{complex operations.}
	\centering
	\begin{tabular}{|r|r|r|}
		\hline
		method
		&2D-SD (for $J_{m,n}$)
		&1d-SD (for $J_k$) \\ \hline
		$QR$ Decomp.
		&$M\times M$ and $N\times N$
		& $MN\times MN$   \\ \hline
		complex $\times$
		&   $\frac{MN(\min(M,N)+3)}{2}$     &     $\frac{MN(1+MN)}{2}$  \\ \hline
		complex $+$
		&    $\frac{MN(\min(M,N)+1)}{2}$    &   $\frac{MN(1+MN)}{2}$    \\ \hline
	\end{tabular}
	\label{table:complexity2}
\end{table}

\subsection{Inverse System}\label{AA}
The Iterative Method (IM) was introduced to address distortion resulting from non-ideal interpolation. By defining $G$ as a distortion operator, it is possible to recursively implement IM to compute $G^{-1}$ in order to compensate for its distortion\cite{shamsi2020nonlinear, shamsi2024acceleration, shamsi2023enhancing}. The IM recursive equation is given by $x_{k} =\lambda (x_{0}-G(x_{k-1}))+x_{k-1}$,
where $x_{k}$ represents the estimated signal after $k$ iterations and $\lambda$ is a relaxation parameter.

In soft decoding\cite{xu2013improved}, the following two steps are taken: $d \leftarrow 1-r/\eta$ and $w \leftarrow \lambda(w_0 -G^{\text{soft}}(w,d))+w$,
where, $p_i$ and $q_i$ denote the real and imaginary components of $w_i$, respectively, and $s_i$ is defined as:
\begin{align*}
s_i = \left\{\begin{matrix}
p_i;\text{if } \left | p_i \right | < d\\ 
sign(p_i);\text{otherwise}
\end{matrix}\right. + 1j\times
\left\{\begin{matrix}
q_i;\text{if } \left | q_i \right | < d\\ 
sign(q_i);\text{otherwise}
\end{matrix}.\right.
\end{align*}
\section{Simulation Results}
\label{sect:sim}
\begin{figure}[h]
	\centering
	\begin{subfigure}[h]{.5\textwidth}
		\includegraphics[trim={0 0 0 0}, scale=0.25]{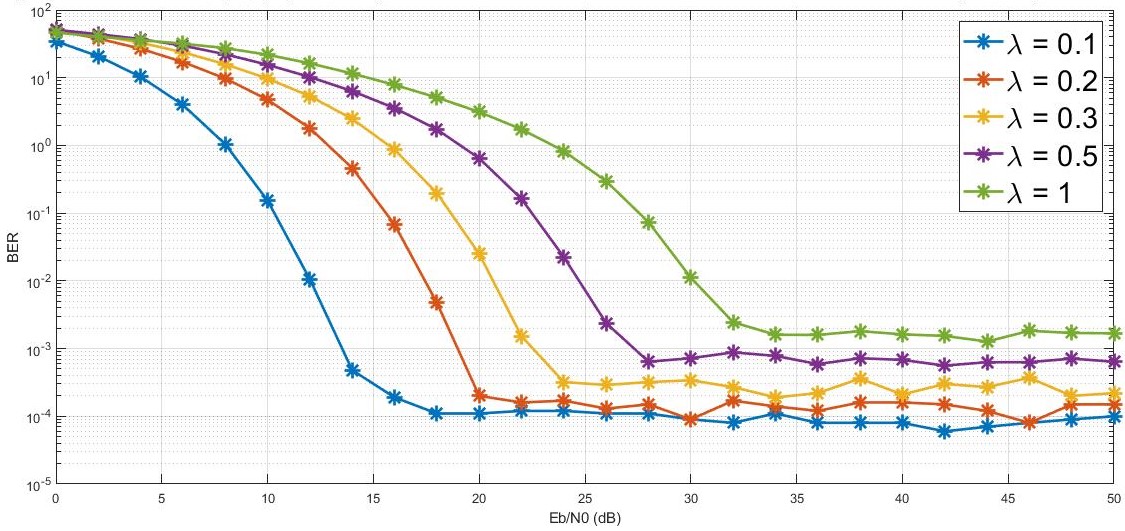}
		\caption{ $(M,N)=(16,16)$, $(\alpha, \beta)= (0.9, 0.9)$ , $\eta=23.5\%$.}
		
		\label{fig:overloaded_DD_11}
	\end{subfigure}%
	\hfill
	\begin{subfigure}[h]{.5\textwidth}
		\includegraphics[trim={0 0 0 0}, scale=0.25]{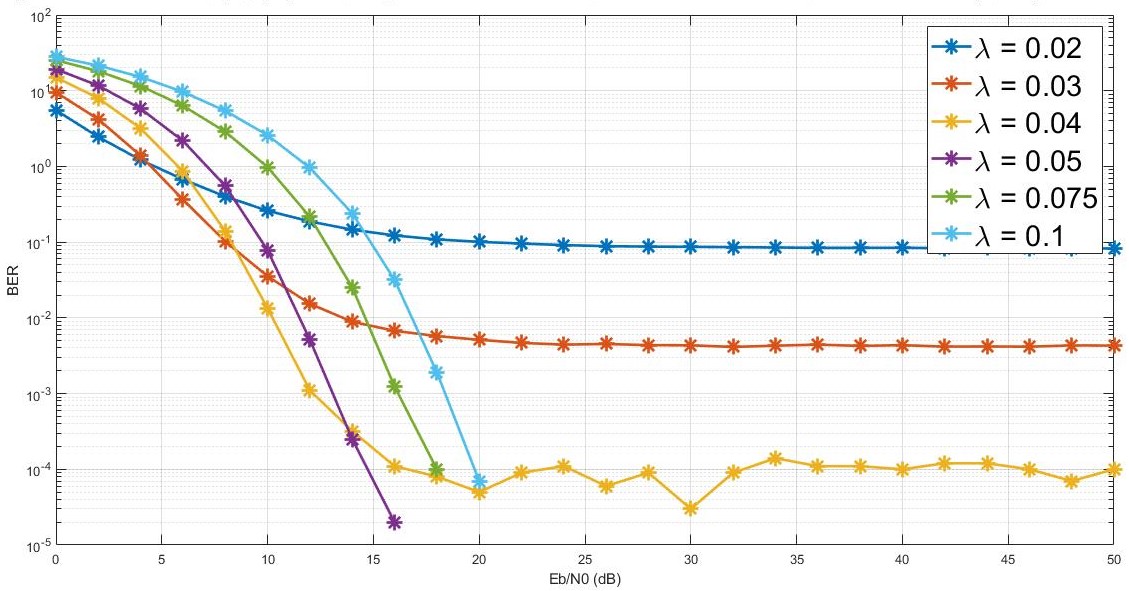}
		\caption{$(M,N)=(8,16)$, $(\alpha, \beta)= (0.85, 0.9)$, $\eta=31\%$.}
		
		\label{fig:overloaded_DD_12}\end{subfigure}%
	\caption{ BER vs EB/N0: NOTFS, AWGN channel, IM and soft decoding (different $\lambda$s).}
	\label{fig:overloaded_DD_1}
\end{figure}

\begin{figure}[h]
	\centering
	\begin{subfigure}[t]{.5\textwidth}
		\includegraphics[trim={0 0 0 0}, scale=0.25]{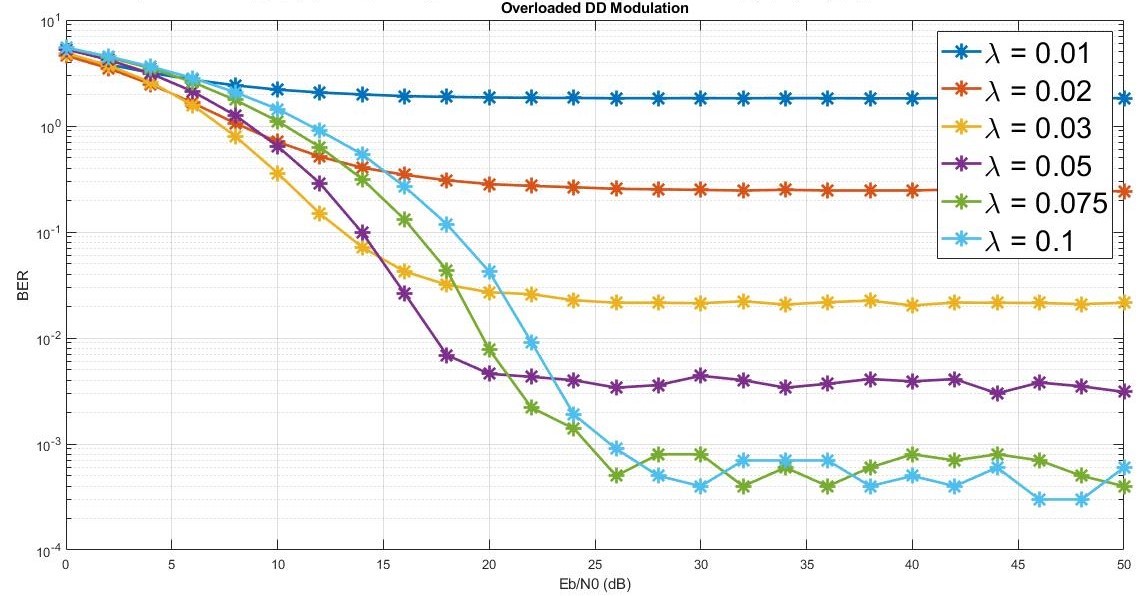}
		\caption{IM: 75 iterations.}
	\end{subfigure}%
	\hfill
	\begin{subfigure}[t]{.5\textwidth}
		\includegraphics[trim={0 0 0 0}, scale=0.25]{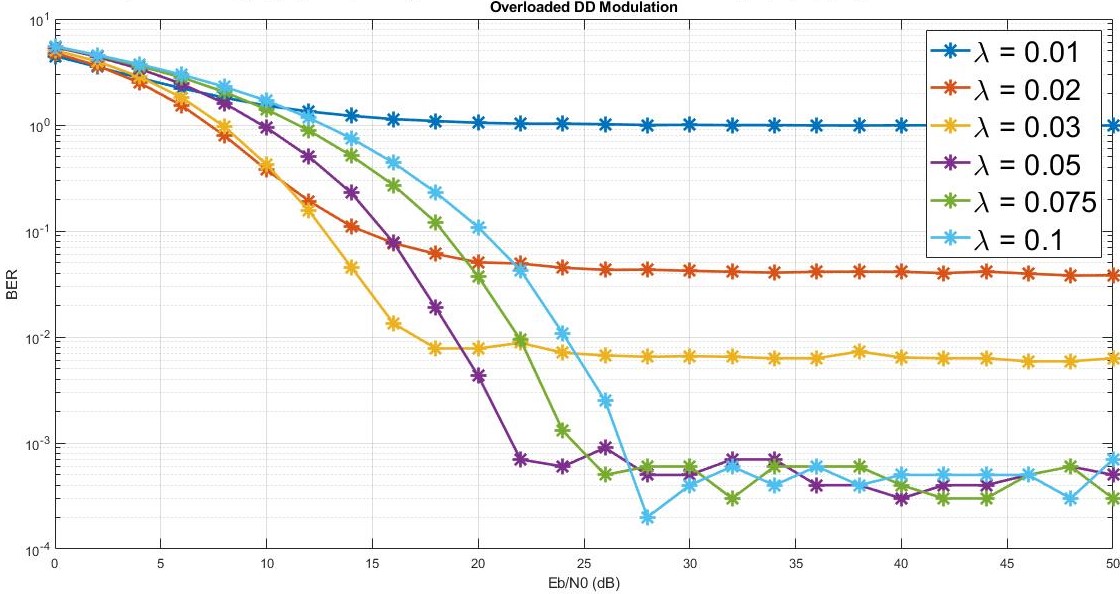}
		\caption{IM: 100 iterations.}
	\end{subfigure}%
	
	\caption{ BER vs EB/N0: NOTFS, AWGN channel, IM and soft decoding ($\lambda$):
		 $(M, N)=(4,4)$,  	$(\alpha, \beta)= (0.675, 0.675)$ equivalent to $\eta=119.5\%$.
	}
	\label{fig:overloaded_DD_4}
\end{figure}

\begin{figure}[h]
	\centering
	
	\begin{subfigure}[t]{0.5\textwidth}
		\includegraphics[trim={0 0 0 0}, scale=0.25]{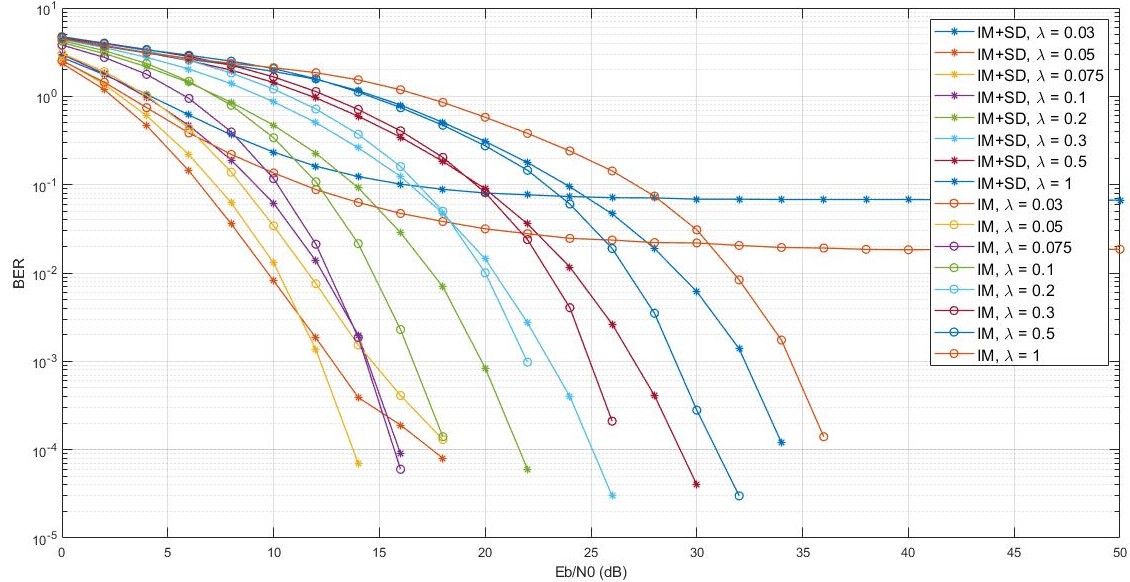}
		\caption{$\alpha=\beta=0.8$, $\eta=56\%$, IM with 30 iterations.}
		\label{fig:overloaded_DD_SD_21}
	\end{subfigure}%
	\hfill
	\begin{subfigure}[t]{0.5\textwidth}
		\includegraphics[trim={0 0 0 0}, scale=0.25]{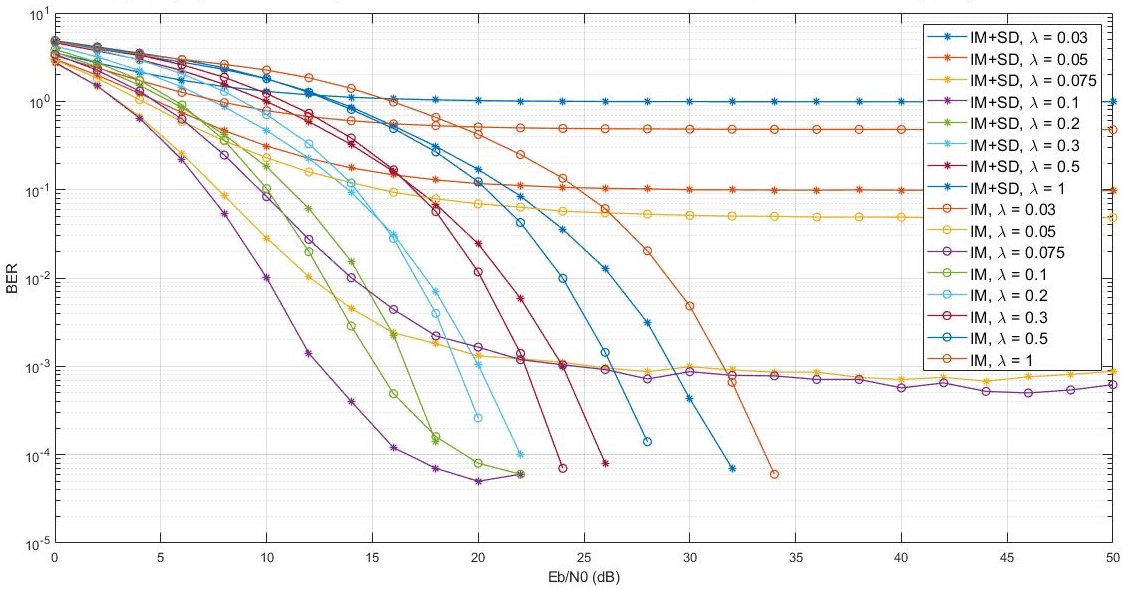}
		\caption{$\alpha=\beta=0.775$, $\eta=66.5\%$`with 20 iterations.}
		\label{fig:overloaded_DD_SD_22}
	\end{subfigure}%
	
	\caption{ BER vs EB/N0: NOTFS, AWGN channel, 2D-SD, initial estimation: IM and soft decoding (different $\lambda$'s):
	 $(M, N)=(4,4)$.
	}
	\label{fig:overloaded_DD_SD_2}
\end{figure}

In this section, we present an assessment of the performance of the proposed modulation techniques and detection algorithms through simulation results. In order to achieve a better understanding, we define the overloading factor as $\eta \triangleq \frac{1}{\alpha.\beta}-1$. The IM with soft decoding is employed to detect the received NOTFS signals in an AWGN channel. The proposed approach shows promising performance by achieving a low BER ($10^{-4}-10^{-5}$) under $30\%$ overloading, as demonstrated by the results shown in Fig. \ref{fig:overloaded_DD_1}
Furthermore, Fig. \ref{fig:overloaded_DD_4} illustrates that the system can attain an acceptable level of performance even with super overloading in small frames. It should be noted that by increasing the number of iterations, the system's performance can be further enhanced.
Through the adoption of the proposed 2-d SD methodology, it is evident that the system's performance can be further improved. As presented in Fig. \ref{fig:overloaded_DD_SD_2}, low values of BER can be achieved even under a high degree of overloading, specifically at $66\%$.
\section{Conclusion}
\label{sect:con}
We introduced a novel modulation technique combining OTFS and OFDM benefits while removing orthogonality constraints, using a Time-Frequency Space Transformation (TFST) to derive non-orthogonal bases over the delay-doppler plane. Our method demonstrated higher spectral efficiency and lower latency, maintaining comparable performance to OTFS and OFDM. Simulation results showcased superior performance in various scenarios. Additionally, a two-dimensional sphere decoding implementation was presented, significantly reducing computational complexity compared to one-dimensional decoding. This technique holds potential as a standard for high-mobility communication systems like 6G, emphasizing vital attributes of low latency and high spectral efficiency.

\bibliographystyle{IEEEtran}
\bibliography{IEEEexample.bib}

\begin{thebibliography}{1}
\providecommand{\url}[1]{#1}
\csname url@samestyle\endcsname
\providecommand{\newblock}{\relax}
\providecommand{\bibinfo}[2]{#2}
\providecommand{\BIBentrySTDinterwordspacing}{\spaceskip=0pt\relax}
\providecommand{\BIBentryALTinterwordstretchfactor}{4}
\providecommand{\BIBentryALTinterwordspacing}{\spaceskip=\fontdimen2\font plus
\BIBentryALTinterwordstretchfactor\fontdimen3\font minus
  \fontdimen4\font\relax}
\providecommand{\BIBforeignlanguage}[2]{{%
\expandafter\ifx\csname l@#1\endcsname\relax
\typeout{** WARNING: IEEEtran.bst: No hyphenation pattern has been}%
\typeout{** loaded for the language `#1'. Using the pattern for}%
\typeout{** the default language instead.}%
\else
\language=\csname l@#1\endcsname
\fi
#2}}
\providecommand{\BIBdecl}{\relax}
\BIBdecl

\bibitem{shamsi2021flexible}
M.~Shamsi, A.~M. Haghighi, N.~Bagheri, and F.~Marvasti, ``A flexible approach
  to interference cancellation in distributed sensor networks,'' \emph{IEEE
  Communications Letters}, vol.~25, no.~6, pp. 1853--1856, 2021.

\bibitem{shamsi2023distributed}
M.~Shamsi and F.~Marvasti, ``Distributed estimation with partially accessible
  information: An imat approach to lms diffusion,'' \emph{arXiv preprint
  arXiv:2310.09970}, 2023.

\bibitem{monk2016otfs}
A.~Monk, R.~Hadani, M.~Tsatsanis, and S.~Rakib, ``Otfs-orthogonal time
  frequency space,'' \emph{arXiv preprint arXiv:1608.02993}, 2016.

\bibitem{mazo1975faster}
J.~E. Mazo, ``Faster-than-nyquist signaling,'' \emph{The Bell System Technical
  Journal}, vol.~54, no.~8, pp. 1451--1462, 1975.

\bibitem{alishahi2011design}
K.~Alishahi, S.~Dashmiz, P.~Pad, and F.~Marvasti, ``Design of signature
  sequences for overloaded cdma and bounds on the sum capacity with arbitrary
  symbol alphabets,'' \emph{IEEE transactions on information theory}, vol.~58,
  no.~3, pp. 1441--1469, 2011.

\bibitem{xu2013improved}
T.~Xu, R.~C. Grammenos, F.~Marvasti, and I.~Darwazeh, ``An improved fixed
  sphere decoder employing soft decision for the detection of non-orthogonal
  signals,'' \emph{IEEE Communications Letters}, vol.~17, no.~10, pp.
  1964--1967, 2013.

\bibitem{shamsi2023enhancing}
M.~Shamsi and F.~Marvasti, ``Enhancing the sefdm performance in high-doppler
  channels,'' \emph{arXiv preprint arXiv:2309.11774}, 2023.

\bibitem{shamsi2020nonlinear}
M.~Shamsi, M.~Ghandi, and F.~Marvasti, ``A nonlinear acceleration method for
  iterative algorithms,'' \emph{Signal Processing}, vol. 168, p. 107346, 2020.

\bibitem{shamsi2024acceleration}
M.~Shamsi and F.~Marvasti, ``Acceleration algorithms for iterative methods,''
  in \emph{Sampling, Approximation, and Signal Analysis: Harmonic Analysis in
  the Spirit of J. Rowland Higgins}.\hskip 1em plus 0.5em minus 0.4em\relax
  Springer, 2024, pp. 521--552.

\end{thebibliography}

\end{document}